\documentclass{ws-procs975x65}
\usepackage{url}
\begin{document}

\title{Stress energy tensor renormalization for a
  spherically symmetric massive scalar field on a quantum space-time}

\author{Nahuel Barrios}
\address{Instituto de F\'{\i}sica, Facultad de Ingenier\'{\i}a,
J. Herrera y Reissig 565,  11300 Montevideo, Uruguay}

\author{Rodolfo Gambini}
\address{Instituto de F\'{\i}sica, Facultad de Ciencias, 
Igu\'a 4225, esq. Mataojo, 11400 Montevideo, Uruguay}

\author{Jorge Pullin}
\address{
Department of Physics and Astronomy, Louisiana State University,
Baton Rouge, LA 70803-4001}

\begin{abstract}
We consider a massive scalar field living on the recently found exact
quantum space-time corresponding to vacuum spherically symmetric loop
quantum gravity. The discreteness of the quantum space time naturally
regularizes the scalar field, eliminating divergences. However, the
resulting finite theory depends on the details of the micro
physics. We argue that such dependence can be eliminated through a
finite renormalization and discuss its nature. This is an example of
how quantum field theories on quantum space times deal with the issues
of divergences in quantum field theories.
\end{abstract}


\bodymatter
\vspace{2cm}
The solution to the constraint equations of loop quantum gravity in
the context of vacuum spherically symmetric space-times has been found
in closed form in loop quantum gravity \cite{sphericalprl}. The
resulting quantum space-time has a discrete nature. It is
characterized by the ADM mass of the space-time, a one dimensional
spin network given by a graph $g$ and a tower of integers
corresponding to the valences of the spin network $\vec{k}$. The
latter are related to the eigenvalues of the triad in the radial
direction $\hat{E}^x$. This quantity in turn determines the areas of
the spheres of symmetry which are therefore quantized. As a
consequence the spacing of the spin network is bounded below by
$\ell_{\rm Planck}^2/(2r)$. For macroscopic situations the spacing is
therefore very small, in fact sub-Planckian.

Quantum fields have been considered on such a space-time. Hawking
radiation has been derived with small corrections \cite{hawking} and 
the Casimir
effect between two spherical shells has been considered \cite{casimir}. The main
difference between considering a quantum field on a quantum space-time
as opposed to a classical space-time is that the field equations
become discretized and the divergences naturally regulated as was
anticipated in \cite{qsd}. Because
the spacing of the lattice is very fine, the resulting discrete
equations can be well approximated by the equations of the
continuum. However, because the lattice spacing is related to the
Planck scale, quantities that would have diverged in the continuum are
finite but large, leaving an unacceptable imprint of the micro physics
on the macro physics. We argue that such dependence must be eliminated
via a (finite) renormalization and the dependence on the micro physics
can be absorbed in a redefinition of the bare constants of the
theory. In this note we would like to sketch the resulting
calculation. We will operate mostly in the continuum as an
approximation to the correct discrete theory and introduce the
discreteness in key steps where it is relevant.

We start from the action for spherically symmetric gravity with a
possible $1+1$ dimensional cosmological constant,

\begin{equation}
  S = \int dx^0 dr \frac{\sqrt{-g^{(2)}}}{4 G_B}\left(R r^2 -2\Lambda_B\right)
\end{equation}
where $G_B$ and $\Lambda_B$ are the bare values of Newton's constant
and the cosmological constant.

We will consider the stress tensor,
\begin{equation}
  \langle T_{\mu\nu} \rangle = \frac{2}{\sqrt{-g}}\frac{\delta
    W}{\delta g^{\mu\nu}},
\end{equation}
where 
\begin{equation}
  W = -\frac{i}{2} {\rm Tr}\left(\ln\left(-G_F\right)\right),
\end{equation}
with $G_F$ the Feynman propagator for the massive scalar field,
\begin{equation}
  \left(\nabla_\mu \nabla^\mu + m^2 +\xi R\right) G\left(x,x'\right) = 
g^{-1/2}(x) \delta(x-x').
\end{equation}
Since we are interested in the short distance behavior of the Green's
function we can expand the metric in Riemann normal coordinates around
a point $x'$, we can expand the metric as
\begin{equation}
  g_{\mu\nu}=\eta_{\mu\nu} -\frac{1}{3} R_{\mu\alpha\nu\beta}y^\alpha
  y^\beta -\frac{1}{6} R_{\mu\alpha\nu\beta,\gamma} 
y^\alpha  y^\beta   y^\gamma+\ldots,
\end{equation}
with $y=x-x'$,
allowing us to expand the propagator as,
\begin{equation}
  G\left(k\right) = \left(k^2-m^2\right)^{-1}
  -\left(\frac{1}{6}-\xi\right) R \left(k^2-m^2\right)^{-2} + \ldots.
\end{equation}

If we now rescale the propagator $\bar{G}= \left(-g\right)^{1/4} G$
and we go to momentum space, and recall that we are in spherical
symmetry so only the radial and time coordinates are involved, we get,
\begin{eqnarray}
  \bar{G}\left(x,x'\right)&=&\int \frac{d^2k}{\left(2\pi\right)^2}
  \exp\left(-ik_0y^0+i
    k_1 y^1\right)\nonumber\\
&& \times \left[1+a_1\left(x,x'\right)\left(-\frac{\partial}{\partial
        m^2}\right) +a_2\left(x,x'\right)
\left(-\frac{\partial}{\partial
        m^2}\right)^2\right]\frac{1}{k_0^2-k_1^2-m^2},
\end{eqnarray}
with,
\begin{equation}
  a_1\left(x,x'\right) = \left(\frac{1}{6}-\xi\right) R -\frac{1}{2}
\left(\frac{1}{6}-\xi\right) R_{,\alpha} y^\alpha -\frac{1}{3}
a_{\alpha\beta} y^\alpha y^\beta.
\end{equation}
The quantity $a_{\alpha\beta}$ is a geometric expression involving
linear and quadratic terms in the scalar curvature, Ricci and Riemann
tensor. In the generic $3+1$ case the term involving $a_2$ also leads
to divergent corrections that need to be compensated introducing
counterterms quadratic in the curvature. In spherical symmetry one
does not need to consider such term, as we shall see.

To compute the Green's function it is good to use the identity,
\begin{equation}
  \left(k^2-m^2\right)^{-1} = -i \int_0^\infty ds \exp\left(i s\left(k^2-m^2\right)\right),
\end{equation}
which allows to integrate in $k$ and yield,
\begin{equation}
  \bar{G}\left(x,x'\right) =-\frac{i}{4\pi} \int_0^\infty \frac{ds}{s} 
\exp\left(-im^2 s +\frac{\sigma}{2 i s}\right)
\left[1+a_1\left(x,x'\right)is +a_2\left(x,x'\right)\left(is\right)^2\right],
\end{equation}
where $\sigma$ is related to the geodesic distance squared between $x$ and
$x'$, $\sigma=y^2/2$. 

This is how the calculation on a classical background goes.However,
the quantum background introduces a difference. As we argued, the
condition of the quantization of the areas of symmetry leads to an
effective quantization of the radial coordinate with $r_i^2 =\ell_{\rm
  Planck}^2 k_i$ with $i$ the label of the vertex of the spin network
associated with the radial position $r_i$. We will consider the
simplest case of a spin network that is equispaced in normal
coordinates with lattice spacing $\Delta$. This imposes a cutoff in
the radial integral in $k^1$ of $2\pi/\Delta$ as is common on a lattice. Then the
Green's function will take the form,
\begin{eqnarray}
  \bar{G}_\Delta\left(x,x'\right) &=& -\frac{i}{8\pi} \int_0^\infty
  \frac{ds}{s} \exp\left(-im^2 s +\frac{\sigma}{2is}\right)
\left[{\rm erf}\left(\frac{\sqrt{i}}{2}\left(\frac{4\pi s-\Delta
          y^1}{\Delta \sqrt{s}}\right)\right)\nonumber\right.\\
&&\left. -
{\rm
      erf}\left(\frac{-\sqrt{i}}{2}\left(\frac{4\pi s+\Delta
          y^1}{\Delta \sqrt{s}}\right)\right)\right]\nonumber\\
&&\times \left[1+a_1\left(x,x'\right)is +a_2\left(x,x'\right)\left(is\right)^2\right].
\end{eqnarray}
As a result the effective action is finite and takes the form
(eq. 6.35 in \cite{birrelldavies}),
\begin{equation}
  W = \frac{i}{2} \int dx^0 dr \sqrt{-g^{(2)}} \lim_{x'\to x}
  \int_{m^2}^\infty dm^2 \bar{G}_\Delta\left(x,x'\right).
\end{equation}
From here we can identify the effective Lagrangian where we study the
divergence, 
\begin{equation}
  L_{\rm effective}^{\rm div} = -\frac{i}{2} \int_0^\infty 
\frac{ds}{s^2}\frac{\exp\left(-im^2 s\right)}{4\pi} 
{\rm erf}\left(\frac{\sqrt{is}2\pi}{\Delta}\right) \left(1+a_1 i
  s\right).
\label{effective}
\end{equation}

For the particular background quantum state we chose with an
equispaced lattice with invariant distance among vertices of the spin
network given by $\Delta$, the first two terms in the expansion in
powers of $(is)$ would lead to divergent contributions in the limit
$\Delta\to 0$. For a finite, sub-Planckian $\Delta$ they are very
large. They can be renormalized by absorbing them in the bare
cosmological constant and bare Newton constant. The total
gravitational Lagrangian density becomes,
\begin{equation}
  L_{\rm ren} = \left(-g^{(2)}\right)^{1/2} \left[-\left(A
      +\frac{\Lambda_B}{8\pi G_B}\right)+\left(B+\frac{r^2}{16\pi G_B}\right)R\right],
\end{equation}
To compute $A$ and $B$ we need to compute the integrals in the
effective Lagrangian (\ref{effective}).  We have done this in two
different ways: a) Using the ascending power series of the error
function and computing the exact sum of the integrals of each term and
analytically extending to the asymptotic region; b) Using the
asymptotic series. In both cases the result is,
\begin{eqnarray}
  A&=& \frac{\pi}{ \Delta^2}
  +\frac{m^2}{8\pi}\left[1+\ln\left(\frac{16\pi^2}{m^2\Delta^2}\right)\right],\\
  B&=& \frac{1}{24\pi}\ln\left(\frac{4\pi}{m\Delta}\right),
\end{eqnarray}
and since we are taking into account only the spherical mode of the
field, the cosmological constant term that arises is not a spherically
symmetric reduction of four dimensional gravity, but it is the
cosmological constant term one has in $1+1$ dimensional theories. 

For the previous calculations, we have assumed that the geodesic 
distance between points of the lattice, which determines the possible
values of $k$, is $\Delta$ and is constant for all points in the
radial direction. This is not really required and the theory allows
variable spacing. However, if one wishes to reabsorb the dependence on
the microscopic states, this will require counterterms that are
state-dependent or to choose privileged states to renormalize. An
intriguing possibility would be to consider conformal gravity where
the size of these distances can be chosen arbitrarily and therefore
one can make the spacing uniform.

Summarizing, the use of quantum field theory in quantum space-time
techniques naturally regularizes the divergences of the field theory,
replacing the divergent terms with terms that are large and depend on
the details of the micro-structure of the background quantum
state. Such dependence can be removed by a (finite) renormalization that
absorbs the dependence on the micro-structure in the coupling
constants of the theory, in this particular model the cosmological
constant and Newton's constant.

This work was supported in part by grant NSF-PHY-1305000, ANII
FCE-1-2014-1-103974, funds of the Hearne Institute for Theoretical
Physics, CCT-LSU and Pedeciba.


\begin{thebibliography}{99}
\bibitem{sphericalprl} R.~Gambini and J.~Pullin,
  Phys.\ Rev.\ Lett.\  {\bf 110}, no. 21, 211301 (2013)
  [arXiv:1302.5265 [gr-qc]];
  R.~Gambini, J.~Olmedo and J.~Pullin,
  Class.\ Quant.\ Grav.\  {\bf 31}, 095009 (2014)
  [arXiv:1310.5996 [gr-qc]].
\bibitem{hawking}  R.~Gambini and J.~Pullin,
  Class.\ Quant.\ Grav.\  {\bf 31}, 115003 (2014)
  [arXiv:1312.3595 [gr-qc]].
\bibitem{casimir}
 R.~Gambini, J.~Olmedo and J.~Pullin,
  Class.\ Quant.\ Grav.\  {\bf 32}, no. 11, 115002 (2015)
  [arXiv:1410.4479 [gr-qc]].
\bibitem{qsd}  T.~Thiemann,
  Class.\ Quant.\ Grav.\  {\bf 15}, 1281 (1998)
  [gr-qc/9705019].
\bibitem{birrelldavies}
N. D. Birrell, P. C. W. Davies, ``Quantum fields in curved space'',
Cambridge University Press, Cambridge, UK (1984).

\end{thebibliography}
\end{document}